\documentclass[doublecol]{epl2}
\usepackage{textcomp}
\usepackage{amsmath}
\usepackage{amssymb}
\usepackage{graphicx} 
\newcommand{\nep}{\textrm{e}}

\newcommand{\calF}{\mathcal F}
\newcommand{\calL}{\mathcal L}

\begin{document}

\title{Loschmidt echo and dynamical fidelity in periodically driven quantum systems}

\author{Shraddha Sharma\inst{1}\and Angelo Russomanno\inst{2,3}\and Giuseppe E. Santoro\inst{2,3,4}\and Amit Dutta\inst{1}}
\shortauthor{S. Sharma \etal}
\institute{
\inst{1} Department of Physics, Indian Institute of Technology Kanpur, Kanpur 208016, India\\
\inst{2} SISSA, Via Bonomea 265, I-34136 Trieste, Italy\\
\inst{3} CNR-IOM Democritos National Simulation Center, Via Bonomea 265, I-34136 Trieste, Italy\\
\inst{4} International Centre for Theoretical Physics (ICTP), P.O.Box 586, I-34014 Trieste, Italy
}

\abstract{
We study the dynamical fidelity $\calF (t)$ and the Loschmidt echo $\calL (t)$, following a periodic driving of the transverse magnetic field 
of a quantum Ising chain (back and forth across the quantum critical point) by calculating the overlap between the initial ground state 
and the state reached after $n$ periods $\tau$.
We show that $[\log{\calF}(n\tau)]/L$ (the logarithm of the fidelity per-site) reaches a steady value in the asymptotic limit $n\to \infty$, 
and we derive an exact analytical expression for this quantity.   
{Remarkably, the steady state value of $[\log{\calF}(n\tau\to \infty)]/L$ shows memory of non-trivial phase information
which is instead hidden in the case of thermodynamic quantities; 
this conclusion, moreover, is not restricted to 1-dimensional models.}
}


\maketitle

The dynamical evolution of closed quantum systems has been recently upgraded from a rather 
academic question to a very intense field of research, due to many experimental advances,
notably cold atom systems \cite{Bloch_RMP08} and femtosecond resolved spectroscopies \cite{Stolow04}.
Many studies, in particular, have focused on different non-equilibrium issues, particularly
when the quantum system is driven out of equilibrium either by a sudden quench or 
by a slow adiabatic change of the Hamiltonian
(the so-called quantum annealing {\it alias} adiabatic quantum computation \cite{Farhi_SCI01,Kadowaki_PRE98,Santoro_SCI02,Das_Chakrabarti:book,Santoro_JPA06}). 
In parallel, there have been numerous studies on the so-called Kibble-Zurek scaling
\cite{kibble76,zurek96,Zurek_PRL05,polkovnikov05,cherng06,mukherjee07} 
of the defect density generated in the final state reached by driving a system across a quantum critical point (QCP) \cite{Sachdev:book}: 
For reviews, see Refs.~\cite{dutta10,polkovnikov10,dziarmaga10}. 
Many works have also explored the connection between quantum phase transitions, 
quantum information \cite{amico08} and quantum critical dynamics \cite{dutta10}. 
Two important measures which show distinct behaviour close to a QCP happen to be the 
Loschmidt echo \cite{quan06,sharma12} and the ground state quantum fidelity \cite{zanardi06}.
Especially the former has been studied extensively in recent years in connection to the dynamics of 
decoherence \cite{damski11,mukherjee12,nag12}, and the work statistics \cite{Silva_PRL08,Pietro_PRE}. 

Considerably less studied is the case of a {\em time-periodic} driving, in which a closed quantum system is 
periodically driven back-and-forth across a QCP \cite{Mucco_JSM09,Arnab_PRB10}. 
Very recently, it has been argued that periodically driven closed quantum systems possessing an absolutely continuous
Floquet spectrum are likely to display a ``periodic steady state'' \cite{Russomanno_PRL12,Russomanno_JSTAT13}, 
i.e, a periodic long-time dynamics for most physical observables, an effect that has been explicitly 
demonstrated for an Ising chain with a periodically driven transverse field.

In this paper we consider the Loschmidt echo \cite{quan06,sharma12,lecho_sch} 
upon a periodic driving of a closed quantum system. 
The Loschmidt echo \cite{lecho_sch} is defined as:
\begin{equation} \label{echo_def:eqn}
{\calL}(t) \equiv \left| \langle \Psi_0 | \hat{U}_0(0,t) \hat{U}(t,0) | \Psi_0\rangle \right|^2
\end{equation} 
where $\hat{U}_0$ and $\hat{U}$ are unitary evolution operators associated to two different 
Hamiltonians $\hat{H}_0(t)$ and $\hat{H}(t)$, and $|\Psi_0\rangle$ is some initial state. 
In the particular case in which $\hat{H}_0$ is time independent and $|\Psi_0\rangle$ 
is its ground state, then the Loschmidt echo is equivalently expressed as:
\begin{equation} \label{fidelity_def:eqn}
{\calL}(t) = \left| \langle \Psi_0 | \hat{U}(t,0) | \Psi_0 \rangle \right|^2 = 
\left| \langle \Psi_0 | \Psi(t) \rangle \right|^2  \equiv {\calF}(t) \;,
\end{equation} 
where the right-hand side might be viewed as a dynamical {\em fidelity}, i.e., the squared overlap
between the initial state and the time-evolved state $|\Psi(t)\rangle=\hat{U}(t,0)|\Psi_0\rangle$,
{\it alias} the return probability to the initial state $|\Psi_0\rangle$,
or the amplitude $P_0$ of the zero-work delta peak, $P_0\delta(W)$,
in the work distribution $P(W)$ \cite{Silva_PRL08,Pietro_PRE}.
We stress that this is different from the conventional ground-state fidelity \cite{zanardi06}  
$\left| \langle \Psi_0 | \Psi_1 \rangle \right|^2$, $|\Psi_{0(1)}\rangle$ being the ground-states of the 
Hamiltonian with two different sets of parameters: If the system repeatedly crosses a QCP during the evolution, 
the defects generated in the process manifest themselves in the dynamical fidelity ${\mathcal F}(t)$.

The key question we will address is whether $\calF(t)$, while not being a standard thermodynamical observable,
would still tend to reach a large-$t$ steady state value \cite{Russomanno_PRL12} when the Hamiltonian 
$\hat{H}(t)$ is time-periodic, $\hat{H}(t+\tau)=\hat{H}(t)$. 
We will show here that the answer is positive for the explicit case we have studied in detail, i.e.,
a periodically driven quantum Ising chain, and we believe that this result extends to all the cases
in which a periodic steady state is found for ordinary observables \cite{Russomanno_PRL12}.
Moreover, and this is the key result of our study, the off-diagonal matrix elements whose 
long-time average is responsible for the periodic steady state of thermodynamical observables play here,
for the dynamical fidelity, an important role: the correct long-time result,
for which we derive an exact analytical expression, does not follow from a decohered density matrix \cite{cherng06}.

In the following, we will concentrate on a simple model where quasi-analytic 
information can be extracted on the quantities of interest, namely the quantum Ising chain in transverse field.
The Hamiltonian of the system is:
\begin{equation}  \label{h11}
\hat{H}(t) =-\frac{1}{2}\sum_{j=1}^{L}\left(J\sigma_j^z\sigma_{j+1}^z + h \sigma_j^x\right)  
                  -\frac{v(t)}{2} \sum_{j=1}^{L_S} \sigma_j^x  \;.
\end{equation}
Here, the $\sigma^{x,z}_j$ are spin-1/2 Pauli matrices at site $j$ for a chain of length $L$ 
with periodic boundary conditions (PBC) $\sigma^{x,z}_{L+1}=\sigma^{x,z}_1$, and
$J$ is the standard longitudinal Ising coupling ($J=1$ in the following). As for the
transverse field terms, involving $\sigma^x_j$,  we allow a uniform piece, $h$, as
well as a time-dependent one, $v(t)$, acting only on a subchain of length $L_S$. 
We will take the time-dependent term $v(t)$ to be periodic, with a single harmonic, i.e., 
$v(t)=A\cos(\omega_0 t+\varphi_0)$.
In the equilibrium case with a homogeneous transverse field (i.e., $A=0$) 
the model has two (mutually dual) gapped phases, a ferromagnetic one ($|h|<1$), 
and a quantum paramagnetic one ($|h|>1$) separated by QCPs at $|h_c|=1$ at zero temperature.
When $A>0$, the transverse field starts to oscillate periodically for $t\ge 0$ around the uniform value $h$, in a region of size $L_S$.
In the following we will concentrate on the case in which $h=h_c$ (i.e., we perturb around the critical Ising Hamiltonian) 
and $L_S=L$ (i.e., the periodic driving acts on the whole chain) for which translational invariance can be exploited.
The translationally non-invariant case can be dealt with using similar techniques, 
as explained in Ref.~\cite{Russomanno_PRL12,Russomanno_JSTAT13}.
Several questions can be addressed when $L_S<L$, in particular the expected markedly different behaviour of the
case in which $L_S={\rm const}$ as $L\to \infty$ (perturbing a finite region inside a very large system, which will act as
``reservoir''), from the case in which $L_S/L={\rm const}$ as $L\to \infty$ (perturbing an extensive region).
The case of a single-site driving, $L_S=1$, is evidently connected with previous decoherence studies, usually 
concerned about sudden quench cases \cite{rossini07}. 
Also, there exist questions regarding the statistics of the work \cite{Silva_PRL08,Pietro_PRE} performed during the driving.
These issues treated in the future, while here we focus on the case $L_S=L$, $h=h_c$, $A=1$ and $\varphi_0=0$.

We shall now proceed to discuss what happens when a periodic driving is performed for a certain number $n$ of
periods $\tau=2\pi/\omega_0$, by looking at the echo/fidelity stroboscopically at time $t=n\tau$, in a wide
range of frequencies $\omega_0$. 
Quan and Zurek \cite{Zurek_NJP10} have performed a similar study following a linear quenching of the transverse field, 
forward and backward, crossing the critical point twice and studying the interference of the phases thus accumulated 
in the wave function, as reflected in the fidelity.  
With a periodic driving scheme we can study the dynamical fidelity 
following successive passages through the QCP and explore its behaviour as a function of $n$
for the  entire range of the driving frequency $\omega_0$, including the extreme adiabatic ($\omega_0 \to 0$) and 
{high-frequency} ($\omega_0 \to \infty$) limits. 
As we shall discuss below, in the limit $n \to \infty$, coherence effects show up prominently for small frequencies, 
whereas in the opposite limit $\omega_0\to \infty$ there is a sort of partial loss of coherence, 
though the dynamics remains fully coherent.

By going to Jordan-Wigner spinless fermions \cite{lieb61} and transforming to $k$-space, one can rewrite 
$\hat{H}(t)$ as a sum of two-level systems (see Refs.~\cite{mukherjee07, Russomanno_PRL12, Russomanno_JSTAT13} for details).
Using the basis states $|0 \rangle$ and $|k,-k \rangle$ for each mode $k$, where $|0 \rangle$ ($|k,-k \rangle$) denotes state with no (two) fermion(s), respectively, the two-level Hamiltonian could be cast into the form
$\hat{H}_k(t) =(-h(t) + \cos k) \, \hat{\sigma}_z + (\sin k) \, \hat{\sigma}_y$ with $\hat{\sigma}_{y,z}$ being standard Pauli matrices 
and $h(t) =1 +\cos \omega_0 t$.
The state of the system can hence be factorized as $|\Psi(t)\rangle = \prod_{k>0} |\psi_k (t)\rangle$ 
with $k =(2p + 1)\pi/L$ and $p=0, \cdots, (L/2 -1)$, as appropriate for fermionic antiperiodic boundary conditions 
(we assume $L$ to be a multiple of $4$). 
Moreover, using the Floquet theory~\cite{Shirley_PR65,Grifoni_PR98} we can re-express each $k$-th component of the state 
$|\psi_k (t )\rangle$ as~\cite{Russomanno_PRL12}:
\begin{equation} \label{psik_t:eqn}
|\psi_k(t)\rangle =r_k^+ \nep^{-i\mu_k t} |\phi_k^+(t)\rangle+r_k^- \nep^{i\mu_k t} |\phi_k^-(t)\rangle \;,
\end{equation}
where $|\phi_k^\pm (t)\rangle$ are the (time-periodic) Floquet modes and $\pm\mu_k$ the corresponding Floquet quasi-energies, 
while the overlap factors $r_k^\pm = \langle\phi_k^{\pm} (0)|\psi_k (0)\rangle$ carry information about the initial state.
Since $|\psi_k (n\tau)\rangle = r_k^+ \nep^{-i\mu_k n\tau} |\phi_k^+(0)\rangle + r_k^- \nep^{i\mu_k n\tau}|\phi_k (0)\rangle$, 
one readily arrives at the following expression for the fidelity after $n$ complete oscillations of the driving field:
\begin{equation} \label{fidelity-ising:eqn}
{\calF}(n\tau) = \nep^{\sum_{k>0}\log |z_n(k)|^2 }  \;,
\end{equation}
where $z_n(k) = |r_k^+|^2\nep^{-i\mu_k n\tau} + |r_k^-|^2\nep^{i\mu_k n\tau}$.
For very large $L$, transforming the sum over $k$ into an integral, one easily establishes that ${\calF}(n\tau)\sim \nep^{Lg_n}$,
where the logarithm of the fidelity (per-site) is well defined for $L\to\infty$ and given by:
\begin{equation} \label{G-fidelity-ising:eqn}
g_n(\omega_0) = \int_{0}^{\pi}\! \frac{dk}{2\pi} \log{|z_n(k)|^2} \;.
\end{equation}
All the Floquet-related quantities appearing in $z_n(k)$ (the overlaps $r_k^{\pm}$ and the quasi-energies $\mu_k$)
are functions of the driving frequency $\omega_0$, a dependence that we have explicitly indicated only in $g_n(\omega_0)$.
For $n=0$, $|z_0(k)|^2=(|r_k^+|^2 + |r_k^-|^2)^2=1$, hence $g_0=0$ and ${\calF}(0)=1$, as expected.
For $n>0$, $|z_n(k)|^2 = |r_k^+|^4 + |r_k^-|^4 + 2 |r_k^+|^2 |r_k^-|^2 \cos(2\mu_{k}n\tau )$ and the
phase-factors related to the Floquet quasi-energies $\mu_k$ will play a role.

\begin{figure}
\begin{center}
      \includegraphics[width=8cm]{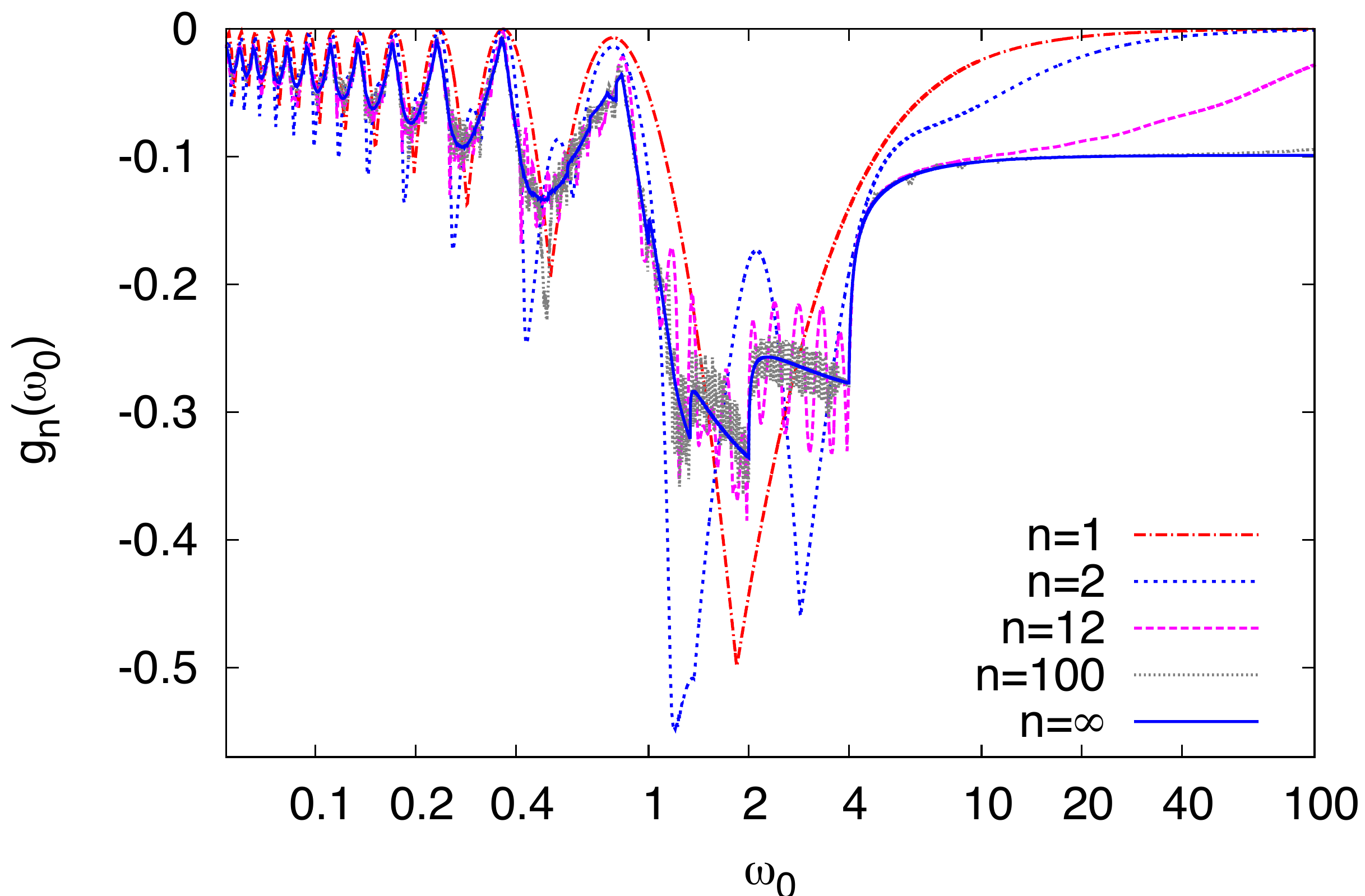}
\end{center}
\caption{$g_n=[\log{\calF (n\tau)}]/L$, the logarithm of the fidelity per-site, as a function of the frequency $\omega_0$
of the driving field $v(t)=\cos{(\omega_0 t)}$ for $n=1, 2, 12, 100$. 
The value of $g_{\infty}$, Eq.~\eqref{G-fidelity-ising-3:eqn}, is also shown.
{Notice how, for any $n$ finite, $g_n(\omega_0)\to 0$ for $\omega_0\to \infty$ (for $n=100$ such a rise towards $0$ is
visible right before $\omega_0\sim 100$).}
}
\label{fidelity:fig}
\end{figure}
\begin{figure}
\begin{center}
      \includegraphics[width=8cm]{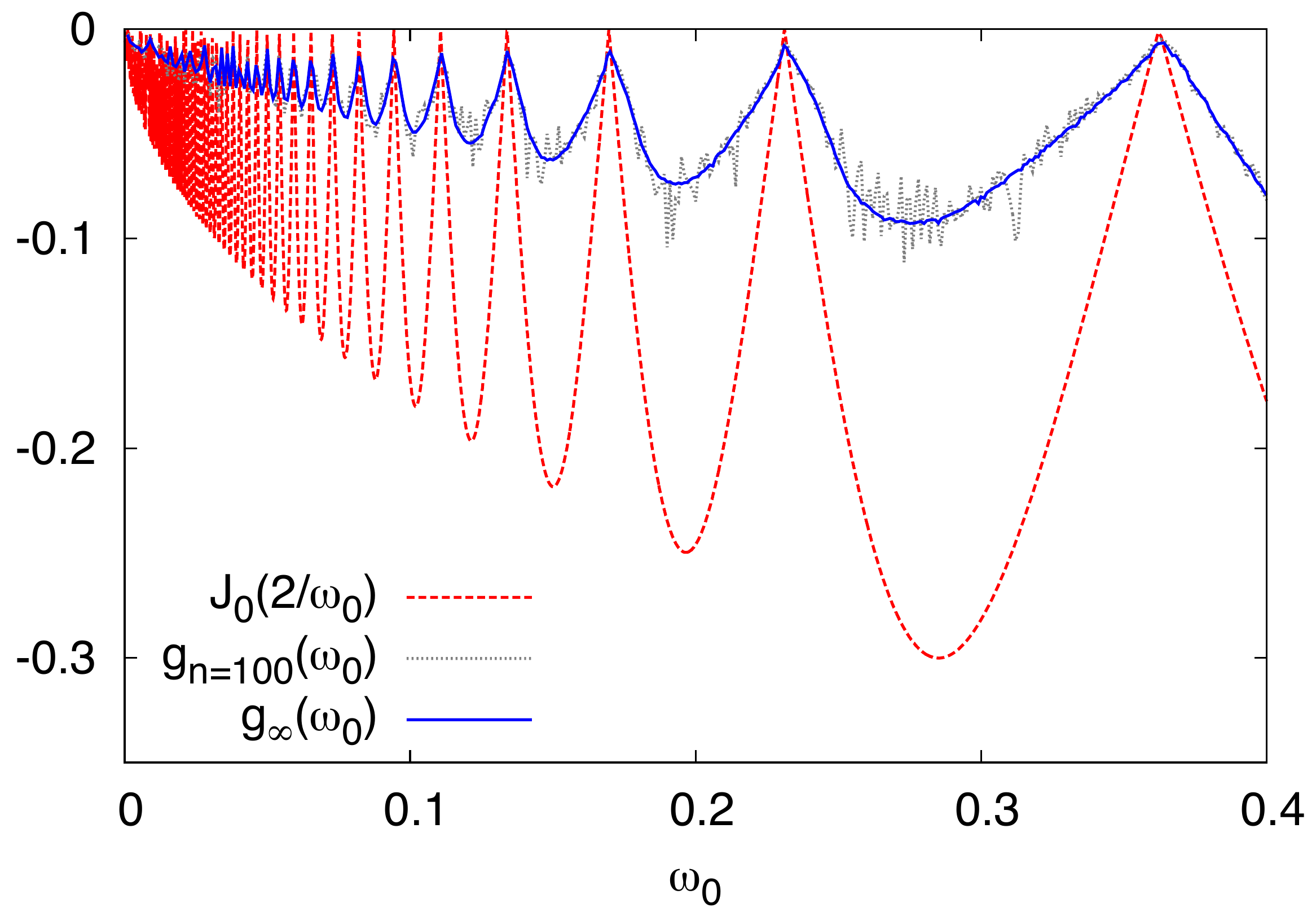}
\end{center}
\caption{Detail of the low-frequency region, showing that the peaks of $g_n(\omega_0)$  (where $g_n\approx 0$, but finite) 
closely match with the zeros of $|J_0(2/\omega_0)|$ for small $\omega_0$, all the way to the limit $n=\infty$. 
(The very small-$\omega_0$ region suffers from numerical accuracy in sampling of the data.)}
\label{bessel:fig}
\end{figure}

%
Fig.~\ref{fidelity:fig} shows the behaviour of $g_n(\omega_0)$ versus the driving frequency $\omega_0$ 
for a few values of $n$. 
Here $n=1$ corresponds to a single oscillation of the field, i.e., a sweep across the QCP at $h_c=1$, 
already showing clear interference effects between the two Kibble-Zurek crossings (back and forth) of the QCP. 
On the other hand, the case $n=2$ shows more marked interference effects, while 
$n=12$, and even more so $n=100$, display a quite rich structure of peaks and dips as a function of $\omega_0$ 
for $\omega_0<4$. 
(The value $4$ marks, in our units, the upper limit of the natural frequencies of the unperturbed critical Ising model.) 
We notice, on the contrary, a rather smooth behaviour for $\omega_0>4$, with 
$g_n(\omega_0)\to 0$ for $\omega_0\to \infty$: the larger the value of $n$, 
the larger will be the typical $\omega_0$ beyond which $g_n(\omega_0)$ rises towards $0$. 
Indeed, the limit $\omega_0\to \infty$, for {\em fixed finite} $n$, involves a kind of {{\em fast driving regime}
in which the system ``does not follow'' the extremely fast oscillations of the field and ``sees'' the average Hamiltonian.} 
This is not hard to justify from Eqs.~\eqref{fidelity-ising:eqn},\eqref{G-fidelity-ising:eqn}:
In the limit $\omega_0\to \infty$, the Floquet quasi-energies $\mu_k$ tend towards the
unperturbed eigenvalues $\epsilon_k$ of the critical Ising model around which we are perturbing;
since in that limit $\tau=2\pi/\omega_0\to 0$, the phase-factors $\nep^{\pm i\mu_k n\tau} \to 1$ for every finite $n$, 
and therefore $|z_n(k)|^2\to (|r_k^+|^2 + |r_k^-|^2)^2=1$, which makes 
$\lim_{\omega_0\to\infty} g_n(\omega_0)=0$.
On the contrary, the behaviour for $\omega_0\le 4$ is highly structured, with peaks and dips
at various frequencies $\omega_0$. 
Notably, for the small values of $\omega_0$,  there are sharp peaks where $g_n(\omega_0)\approx 0$ at all frequencies
for which $J_0(2/\omega_0)=0$, where $J_0$ is the zeroth order Bessel function; 
this is a quite common finding in this field, related to the phenomenon of ``coherent destruction of tunnelling'' \cite{Grossmann_PRL91,Grifoni_PR98}. 
We will address this point later, see Fig.~\ref{bessel:fig} and accompanying discussion.  

An obvious question arising from the inspection of Fig.~\ref{fidelity:fig} concerns what happens as $n\to \infty$: 
will $g_n$ tend towards some definite limit, and if so, how can we calculate it?
This question has to do with whether the system will effectively ``synchronize'' with the driving, i.e., if
the stroboscopic observation of $g_n(\omega_0)$ will reveal a {\em steady state} reached for $n\to \infty$.  
The issue of reaching a ``periodic steady state'' when a closed quantum system is periodically driven
has been raised and analyzed in detail, precisely for the model we are considering, in Ref.~\cite{Russomanno_PRL12}.
There, it was shown that extensive physical quantities --- like the energy density, the density of defects generated, and the transverse 
magnetization --- will indeed reach a periodic steady state predicted by the diagonal elements in the Floquet expansion, 
while off-diagonal terms will lead to a vanishing contribution for $n\to \infty$, due to destructive interference effects 
associated to the widely oscillating phase-factors $\nep^{\pm i\mu_k n\tau}$ when integrated over all momentum modes: mathematically, this is a consequence of the 
Riemann-Lesbesgue lemma predicting the vanishing, at large times, of Fourier transforms of sufficiently regular spectral densities. 
This can be rephrased by saying that the pure-state 
$\hat{\rho}_n=|\Psi(n\tau)\rangle\langle \Psi(n\tau) |$ 
can be effectively replaced, when calculating the average of most extensive physical quantities, with
its {\em decohered} part \cite{cherng06}. In the Floquet basis, we would express this as follows: 
\begin{eqnarray} \label{rho-floquet-decoh:eqn}
\hat{\rho}_n &=&
\prod_{k>0}  \left[ \begin{array}{cc}
                              	|r_k^+|^2 					&  \!\! r_k^+ r_k^{-^*} \nep^{-2i\mu_kn\tau} \!\! \\
                               	\!\! r _k^- r_k^{+^*} \nep^{2i\mu_kn\tau} & |r_k^-|^2 \end{array} \right] 
\stackrel{\scriptscriptstyle n\to\infty} {\longrightarrow}\hat{\rho}_{\rm dec} \hspace{2mm} \nonumber \\
\hat{\rho}_{\rm dec} &=&
\prod_{k>0}  \left[ \begin{array}{cc}
				|r_k^+|^2 	&  	0 \\
				0 			& 	|r_k^-|^2 \end{array} \right] \;. \hspace{15mm}
\end{eqnarray}
For the fidelity, however, the highly-oscillating phase factors due to the off-diagonal terms play a 
trickier and more important role. To better appreciate this, let us rewrite $g_n$ as:
\begin{equation} \label{G-fidelity-ising-bis:eqn}
g_n(\omega_0) = \int_{0}^{\pi}\! \frac{dk}{2\pi} \log{\frac{1+q_k \cos(2\mu_{k}n\tau )}{1+q_k}} \;,
\end{equation}
where $q_k\equiv2|r_k^+|^2|r_k^-|^2/(|r_k^+|^4 + |r_k^-|^4)\in [0,1]$. 
For very large $n$, the $\cos(2\mu_{k}n\tau )$ term oscillates rapidly as $k$ runs over the BZ, 
and one would be tempted to conclude that $g_{n\to\infty} \to g_{\rm dec}(\omega_0)=-\int_{0}^{\pi}\! \frac{dk}{2\pi} \log{(1+q_k)}$,
which is the result predicted by taking $g_{\rm dec}= [\log{\langle \Psi(0) | \hat{\rho}_{\rm dec} | \Psi(0) \rangle}]/L$.
This conclusion, however, would be swift and wrong. \textcolor{black}{We could indeed write, using $\log(1+x)\le x$, that:
\begin{eqnarray}  \label{Echo_upperbound:eqn}
g_n(\omega_0) &=& g_{\rm dec}(\omega_0)  + \int_{0}^{\pi}\! \frac{dk}{2\pi} \log{[1+q_k \cos(2\mu_{k}n\tau )]} \nonumber \\
&\le & g_{\rm dec}(\omega_0)  + \int_{0}^{\pi}\! \frac{dk}{2\pi} \; q_k \cos(2\mu_{k}n\tau ) \;,
\end{eqnarray}
where the integral on the right-hand side of the second expression tends to $0$ for $n\to\infty$ 
(by the Riemann-Lesbesgue lemma), but this does not allow us to conclude that 
$\lim_{n\to \infty} g_n = g_{\rm dec}$.  
}
The analysis of this limit is rather intricate in the general case. 
For $\omega_0>4$, however, things are much simpler, because $q_k$ reaches its maximum value, 
$q_k=1$, only at $k=0$, where definitely $\cos(2\mu_{k}n\tau)\neq -1$ and hence there are no convergence 
issues in the expansion of the $\log[1+q_k \cos(2\mu_{k}n\tau)]$.
%
%
By expanding $\log[1+q_k \cos(2\mu_{k}n\tau)]$, one can show that there are $\cos(2\mu_{k}n\tau)$-related 
terms that survive when $n\to \infty$, originating from all even power terms of the expansion, for the simple reason 
that $\cos^{2p}(2\mu_k n\tau)$ (with $p$ integer) does not average to zero even if $n$ is very large and the oscillations are very fast.
For $\omega_0>4$ we can derive the following closed analytical expression for $g_{\infty}$ which, quite amusingly,
fits our numerical data perfectly well even for $\omega_0<4$:
%
\begin{eqnarray} \label{G-fidelity-ising-3:eqn}
%
\lim_{n\to \infty} g_n(\omega_0) &=& g_{\rm dec}(\omega_0) 
-\sum_{p=1}^\infty \frac{(2p-1)!!}{2p(2p)!!} \int_{0}^{\pi}\! \frac{dk}{2\pi} q_k^{2p}\nonumber\\ 
&=& g_{\rm dec}(\omega_0)  - \int_{0}^{\pi}\! \frac{dk}{2\pi} \log{\frac{2}{1+\sqrt{1-q_k^2}}} \nonumber \\
&=& - \int_{0}^{\pi}\! \frac{dk}{2\pi} \log{\frac{2(1+q_k)}{1+\sqrt{1-q_k^2}}} \;.
%
\end{eqnarray}
%
%
%
%
%
{Observe that, for large $\omega_0$, neither the $q_k$ nor the Floquet states on which they depend show a dependence on $\omega_0$.
This is a sign of the \emph{fast driving regime}: when the driving is fast, the system is not able to follow the oscillations of the field
and ``sees'' the average Hamiltonian. 
This leads to the high-frequency plateau visible in Fig.~\ref{fidelity:fig}.}

%

\begin{figure}
\begin{center}
      \includegraphics[width=8cm]{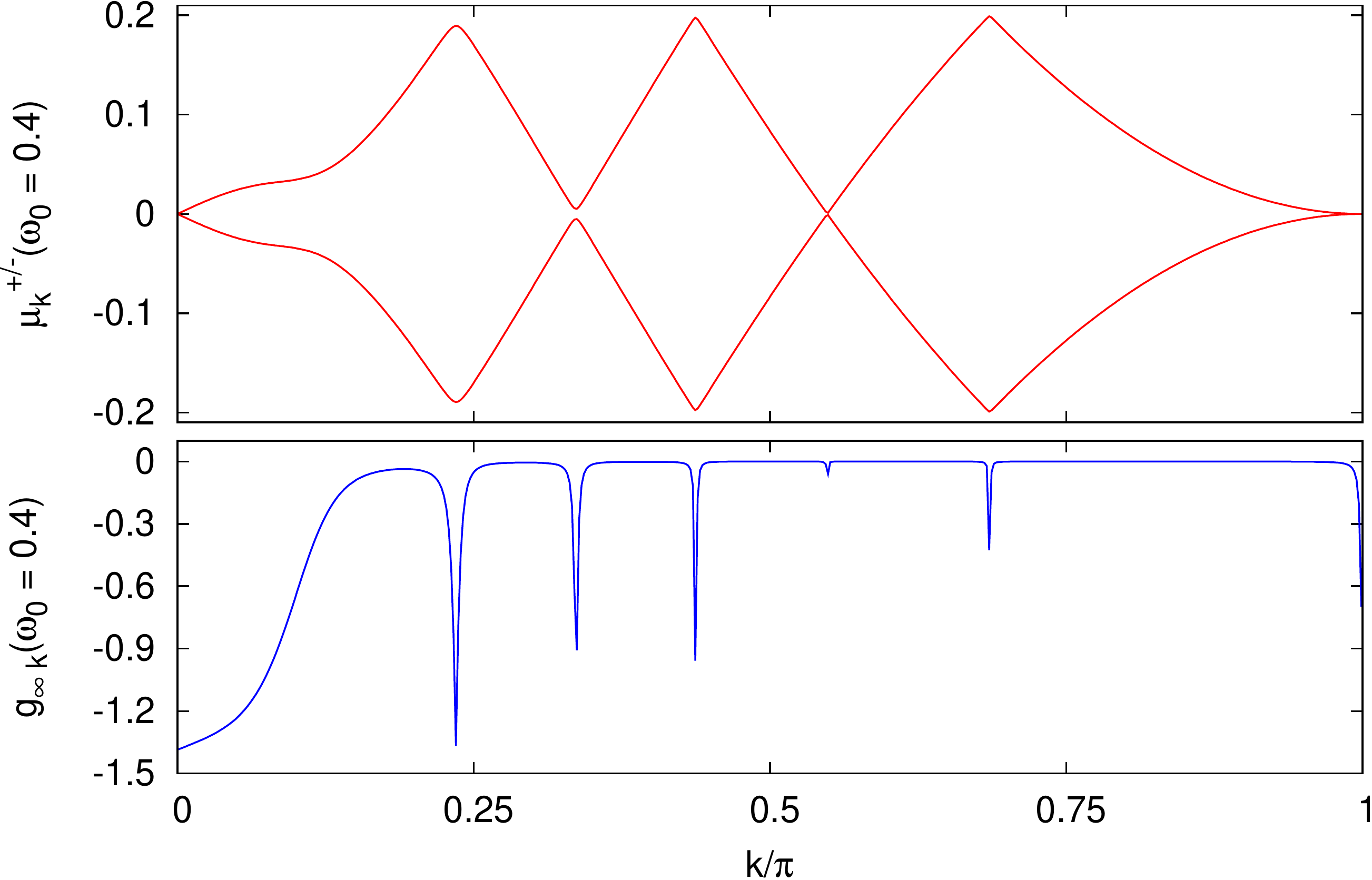}
\end{center}
\caption{The integrand of $g_{\infty}(\omega_0)$ in Eq.~\eqref{G-fidelity-ising-3:eqn} (lower panel) versus $k$ for $\omega_0=0.4$,
together with the Floquet quasi-energies $\pm \mu_k$ (upper panel). The sharp negative features produce dips in 
$g_{\infty}$ (and $g_n$, not shown) at $\omega_0=4/m$ with $m=1,2\cdots $ (here $m=10$), 
when a resonance at $k=\pi$ is about to enter/leave the $k\in[0,\pi]$ integration region.
}
\label{quasi-spec:fig}
\end{figure}

Even in the regime $\omega_0<4$, the overall structure of $g_n(\omega_0)$ for small/intermediate $n$ 
is perfectly predicted by the analytical expression in Eq.~\eqref{G-fidelity-ising-3:eqn}.
In the small-$\omega_0$ regime, there are peaks of $g_n(\omega_0)\sim 0$ (small but finite negative values, really) 
at all the frequencies for which $J_0(2/\omega_0)=0$. 
This result can be understood by looking at the modes close to the critical one at $k=0$, for which $|k|\ll \omega_0$.
Investigating the reduced two-level Hamiltonian in this limit
(with a bias energy between the two states, appearing in the diagonal terms of the reduced Hamiltonian, 
$\epsilon_0=1-\cos k\sim k^2/2$ for $k\to 0$, and the off-diagonal term $\sim k \ll \omega_0$), 
and assuming that the system is initially (at $t=0$) in the state $|k,-k \rangle$, 
the squared amplitude of the state $|0 \rangle$ at time $t$ can be shown to be given by $\sin^2\left[tkJ_0(\frac{2}{\omega_0})\right]$ 
(see Eq.~(7) of Ref.~\cite{Kayanuma_PRA94}). 
Therefore, a coherent destruction of tunneling \cite{Grossmann_PRL91,Grifoni_PR98}
occurs for those values of $\omega_0$ for which $J_0(2/\omega_0)=0$: 
the system sticks to its initial state, resulting in the peaks shown in Fig.~\ref{bessel:fig}.
The regime $\omega_0<4$ also shows clear dips occurring for $\omega_0=4/m$ with $m=1,2,\cdots$, 
due to quasi-degeneracies in the Floquet spectrum \cite{Russomanno_PRL12}. 
As shown in Fig.~\ref{quasi-spec:fig}, the integrand of $g_{\infty}$, for instance, when analyzed versus $k$, shows clear negative 
sharp features at all $k$-points where the Floquet quasi-energies $\pm \mu_k$ become quasi-degenerate, 
either at $0$ or at $\pm \omega_0/2$: these negative sharp features, in particular their crossing of the $k=\pi$ boundary at certain 
frequencies, are in the end responsible for the sharp dips observed in $g_n(\omega_0)$. 

\begin{figure}
\includegraphics[width=8cm]{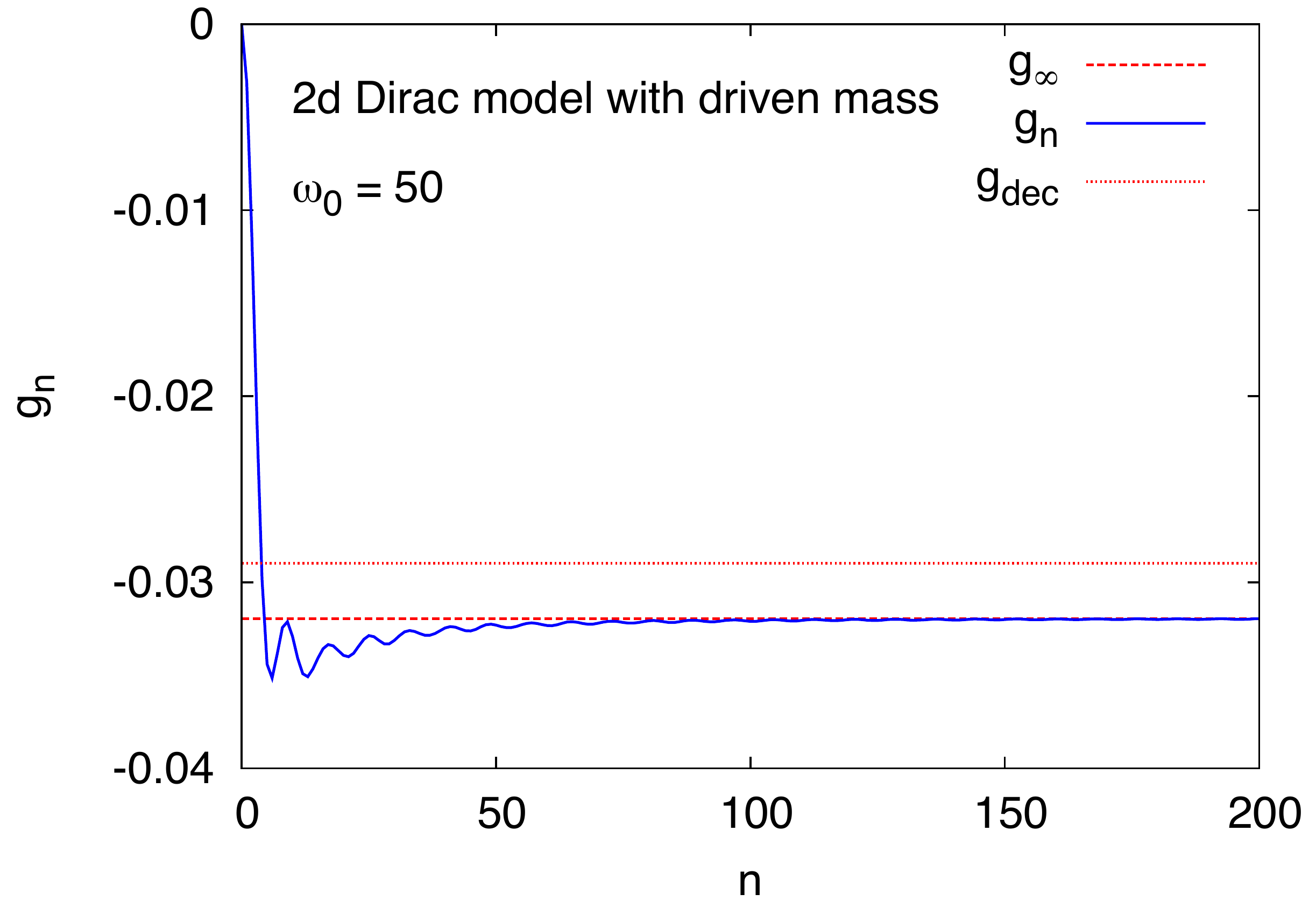}
{\caption{$g_n=[\log{\calF (n\tau)}]/L$, the logarithm of the fidelity per-site, for a two-dimensional massive Dirac Hamiltonian \eqref{Eq:Ham_Massive_Dirac} with a periodic mass driving $m(t)=m_0\cos{(\omega_0 t)}$ versus $n$, for frequency $\omega_0=50$. 
The numerical calculation has been performed by discetizing the Brillouin zone momenta as $k_{x,y}=\frac{2\pi}{L a}n_{x,y}$ 
with $L=1000$ and $n_{x,y}=0 \cdots L-1$. 
Observe that $g_n$ quickly saturates to the analytical limit $g_{\infty}$ for $n\to\infty$, well below  
the value $g_{dec}$ obtained by using the decohered density matrix.}}
\label{fig_dirac}
\end{figure}
%
{A crucial question is how much of this phenomenology depends on the one-dimensional nature of our system, and on the fact that
the present model is integrable. It is difficult to firmly assess the role of integrability, since non-integrable systems are very difficult to tackle:
as argued in Ref.~\cite{Russomanno_PRL12}, we believe however that the crucial ingredient is --- rather than integrability --- 
an absolutely continuous Floquet spectrum.
Concerning the issue of dimensionality, it is simple to provide examples in higher dimensions where the same results can be found. 
Consider, for instance, a two-dimensional massive Dirac Hamiltonian
$\hat{H}_D = \int \! d{\bf x} \; \widehat{\Psi}^{\dagger}({\bf x}) \Big[ m \hat{\sigma}_z -i\hbar v_F (\hat{\sigma}_x \partial_x + \hat{\sigma}_y \partial_y ) \Big] \widehat{\Psi}({\bf x})$,
where $\widehat{\Psi}({\bf x})$ is a two-component spinor field operator describing, for instance, the effective low-energy degrees of
freedom for electrons on a graphene lattice with unequal sublattice potentials around a single Dirac point with Fermi velocity $v_F$ \cite{castroneto09}.
Variations of this model are ubiquitous in the field of topological insulators \cite{Kane_RMP10,Zhang_RMP11}.
The model has also been considered in the context of the ground state fidelity and Loschmidt echo in Ref.~\cite{patel131}. 
It shows a quantum phase transition when the mass $m$ is tuned to zero. 
In momentum space, the Hamiltonian can be written as
\begin{equation} \label{Eq:Ham_Massive_Dirac}
{\mathcal H}_D({\bf k}) = \left[ \begin{array}{cc} 
m & \hbar v_F(k_x-ik_y) \\ 
\hbar v_F(k_x+ik_y) &- m
\end{array} \right] \;,  
\end{equation}
}
{
with a momentum cutoff of $|k_{x,y}|\le \pi/a$. 
Suppose the mass $m$ is periodically driven across the QCP, $m(t)=m_0 \cos\omega_0 t$. 
The dynamical fidelity ${\cal F}(n \tau)$ can then be calculated by factorizing the state in terms of two-component spinor
wavefunctions $|\psi_{\bf k}\rangle$ for each momentum, obtaining a result entirely similar to Eq.~\eqref{G-fidelity-ising-bis:eqn}, i.e.:
\begin{equation} \label{eq_dirac1}
g_n(\omega_0) = \int_{0}^{\frac{\pi}{a}}\! \frac{dk_x dk_y}{(2\pi)^2} \log{\frac{1+q_{\scriptscriptstyle{k_xk_y}} \cos(2\mu_{\scriptscriptstyle{k_xk_y}}n\tau )}{1+q_{\scriptscriptstyle{k_xk_y}}}} \;.
\end{equation}
From this relationship, following an identical line of arguments, one can
show that $g_n$ indeed saturates, in the limit $n \to \infty$, to the analog of the analytic formula $g_\infty$ in
Eq.~\eqref{G-fidelity-ising-3:eqn}.
An instance of this convergence is presented in Fig.~(\ref{fig_dirac}), where we show $g_n$ vs $n$ evaluated
numerically for the Dirac model in Eq.~\eqref{Eq:Ham_Massive_Dirac} when $\omega_0=50$. 
We see that $g_n$ indeed saturates to a steady state value $g_\infty$ which is definitely below the value of $g_{\rm dec}$
obtained from the decohered density matrix. 
This picture holds true for all the frequencies we have checked, showing that also in this 2-dimensional case the 
asymptotic dynamical fidelity retains phase information which is lost in the thermodynamical observables.}

In conclusion, we have studied the dynamical fidelity $\calF(n\tau)$ of a quantum Ising chain following an $n$-period 
sinusoidal driving of the transverse field across the QCP.
We addressed the question of whether $g_n=[\log{\calF(n\tau)}]/L$ saturates to a well defined limit for large $n$, 
when the system reaches a periodic steady state as manifested in extensive thermodynamic quantities \cite{Russomanno_PRL12}. 
The answer is not really obvious: the presence of the logarithm in the expression of $g_n$ forbids a direct application of 
the Riemann-Lesbesgue lemma and complicates the situation.  
Our results confirm that indeed $g_n$ saturates to a steady value, and we were able to derive an exact expression for $g_{\infty}$
which appears to work perfectly well for all values of $\omega_0$, yielding very detailed information on the exact position of peaks and
dips of $g_n$ even at intermediate values of $n$.
However, the surprising result of our analysis is that $g_{\infty}$ {\em cannot} be derived from the fully decohered 
mixed-state density matrix, $\hat{\rho}_{\rm dec}$, where all phase information contained in the off-diagonal 
matrix elements is ``effectively'' lost, as instead possible for the thermodynamic quantities.
Indeed, we have shown that $g_n$ (and $g_{\infty}$) is always strictly smaller then the result derived from $\hat{\rho}_{\rm dec}$.
In the high frequency region ($\omega_0\gg 4$), we find that $g_{\infty}(\omega_0)$ saturates to a finite negative value
{independent of $\omega_0$ showing that the system is unable to follow the oscillations of a too fast driving.
With the example of a Dirac Hamiltonian in two dimensions, we have also shown 
that our main results are not limited to one-dimensional models.}
\acknowledgments
We acknowledge discussions with A. Silva, R. Fazio, M. Fabrizio, P. Smacchia, and E. Tosatti. 
Research at SISSA was supported by MIUR, through PRIN-20087NX9Y7, 
by SNSF, through SINERGIA Project CRSII2 136287\ 1, 
by the EU-Japan Project LEMSUPER, and by the EU FP7 under grant agreement n. 280555. 
AD and SS acknowledge Abdus Salam ICTP, Trieste, where the initial part of the work was done. 
SS acknowledges CSIR, New Delhi, for financial assistance.


\end{document}